\documentclass{jltp}
\usepackage{graphicx}
\title{Threshold effect during  dissolution of\\  $^3$He inclusions
in solid $^4$He}

\author{A.Ganshin, V.Grigor'ev, V.Maidanov, N.Omelaenko,\\ A.Penzev,
 E.Rudavskii, and A.Rybalko}
\address{Laboratory of Quantum Fluids and Solids\\
B.Verkin Institute for Low Temperature Physics \& Engineering \\
310164, Lenin ave 47 Kharkov, Ukraine}

\runninghead{A. Ganshin et al.}
{Threshold effect during dissolution of $^3$He inclusions in solid $^4$He}
\begin{document}
\maketitle
\begin{abstract}
A pressure jump has been found at the onset of the dissolution of bcc
inclusions  in separated solid $^3$He - $^4$He mixture
if the crystal is overheated  above a certain critical
value. This effect can be  explained in the framework
of a multistage dissolution process model.

\end{abstract}

\maketitle

\section{INTRODUCTION}

 The interest in the investigation of the phase separation kinetics of solid
 $^3$He - $^4$He mixtures has been remaining for a long time.
 This interest is
stimulated by the prospect to discover new additional features of the
$^3$He impurities quantum motion in solid helium. In particular, the
diffusion coefficient of impurities during the phase separation can differ
from one measured in NMR experiments in a homogeneous crystal due to
 the influence of a finite concentration gradient and the
distinction of U-processes contribution. However, the
available experimental data on the phase separation kinetics appear to be
badly reproducible and  do not allow one to make any comparisons with
the results obtained in the study of the quantum diffusion.

A good reproducibility of the experimental data was attained quite recently.
\cite{separ,ourfirst}
 The correlation  was established
 between the results obtained in the investigation of the separation
 kinetics and in the NMR experiments in a homogeneous mixture
  \cite{separ}. Nevertheless, it was shown in
\cite{ourfirst,oursecond} that the correlation
 of these data has a more complicated character. The impurity motion
 mechanism during the separation of solid $^3$He - $^4$He mixtures
 remains uncertain. This  calls for continuing
 the investigation and in particular clearing up the bcc phase inclusions
 dissolution kinetics, which has been scarcely studied. The revealed  in
 \cite{ourfirst}
 essential difference between inclusions growth and dissolution
processes adds interest to such investigations.

This work is devoted to a detailed investigation of the $^3$He
inclusions dissolution, which takes place at sharp warming the
separated solid mixture.

\section{KINETICS OF DISSOLUTION OF $^3$He INCLUSIONS IN $^4$He MATRIX}
 The time dependences of the pressure $P$ at a constant volume in a
sample after heating has been measured.
The experimental setup and technique  are discussed in detail
in \cite{ourfirst}.
 The initial concentration of the mixture was
 2.05 \% of $^3$He. The molar volume of the sample before separation
 was 20.44 cm$^3$/mole ($P = 33.4$ bar).
 All the experiments were carried out with the sample after stabilization
of its properties by the many times
repeated cycles of the inclusions growth (at $T\approx 103$ mK)
  and dissolution (at $T\approx 230$ mK) with achievement of equilibrium
at each temperature.

\begin{figure}[p]
\begin{center}
\leavevmode
\includegraphics[scale=0.58,bb = 110 17 500 840,clip]{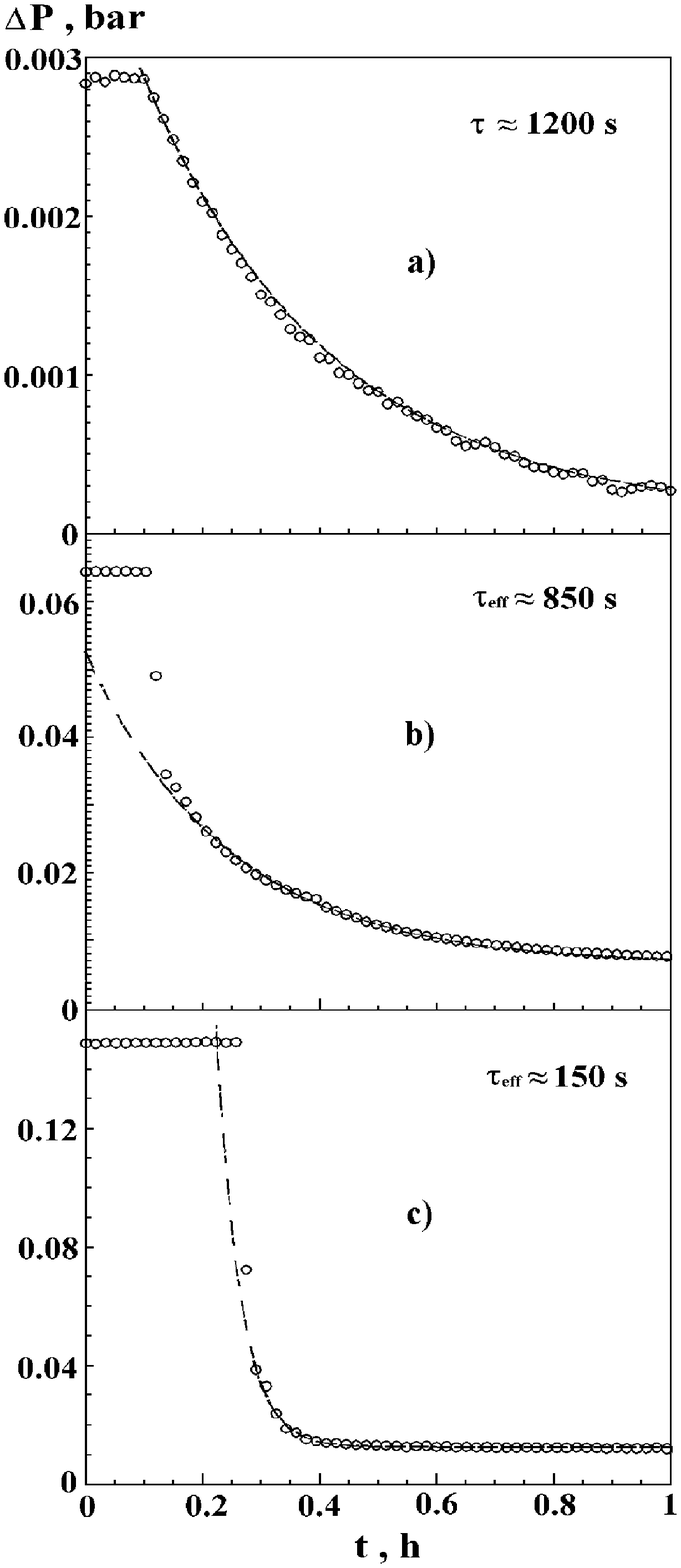}
\caption{ Time dependence of  pressure change in the sample during
 $^3$He inclusion dissolution at
 various temperatures: a) $T_f=110$ mK b) $T_f=150$~mK
c) $T_f=570$ mK. The dashed lines are fits to Eq.(1) with the corresponding
time constants $\tau $ shown in this figure.}

\label{fig1}
\end{center}
\end{figure}

The main set of experiments included  measurements of dependences P(t) at
sharp (within several seconds) heating of the sample from the same initial
temperature $T_i=$103 mK to various final temperatures $T_f=108\div690$ mK.
Results  are partially presented in Fig.\ref{fig1}.
 The shape of these curves is essentially conditioned by the value
of $T_f$.
On small heating
(Fig.\ref{fig1}a) the pressure change $\Delta P$ proved to be exponential
 as in the most of experiments on the separation kinetics of solid
 $^3$He - $^4$He mixtures:
\begin{equation}
\Delta P=\Delta P_0~e^{-t/\tau },
\end{equation}
where $\Delta P_0$ is  difference of equilibrium pressure values
 in the sample at the initial and final temperatures.


At intermediate values of $T_f=127\div230$ mK (Fig.\ref{fig1}b) two different
 parts of the dependence P(t)
are observed: almost vertical part (corresponding time is of the order
of the time of reaching thermal equilibrium) and smoother one described
 by the time constant $\tau_{eff}$, which is of the same order
 as that observed at small $T_f$. The time constant $\tau_{eff}$
 is some effective value obtained at fitting the smooth part of P(t)
by one exponent. In fact, this portion of $P(t)$ can be described by
two exponents  within our experimental accuracy.
And finally, the pressure relaxation occurs very fast at high
temperatures of $T_f$ (Fig.\ref{fig1}c).
\begin{figure}[t]
\begin{center}
\leavevmode
\includegraphics[scale=0.5,bb = 30 170 590 680,clip]{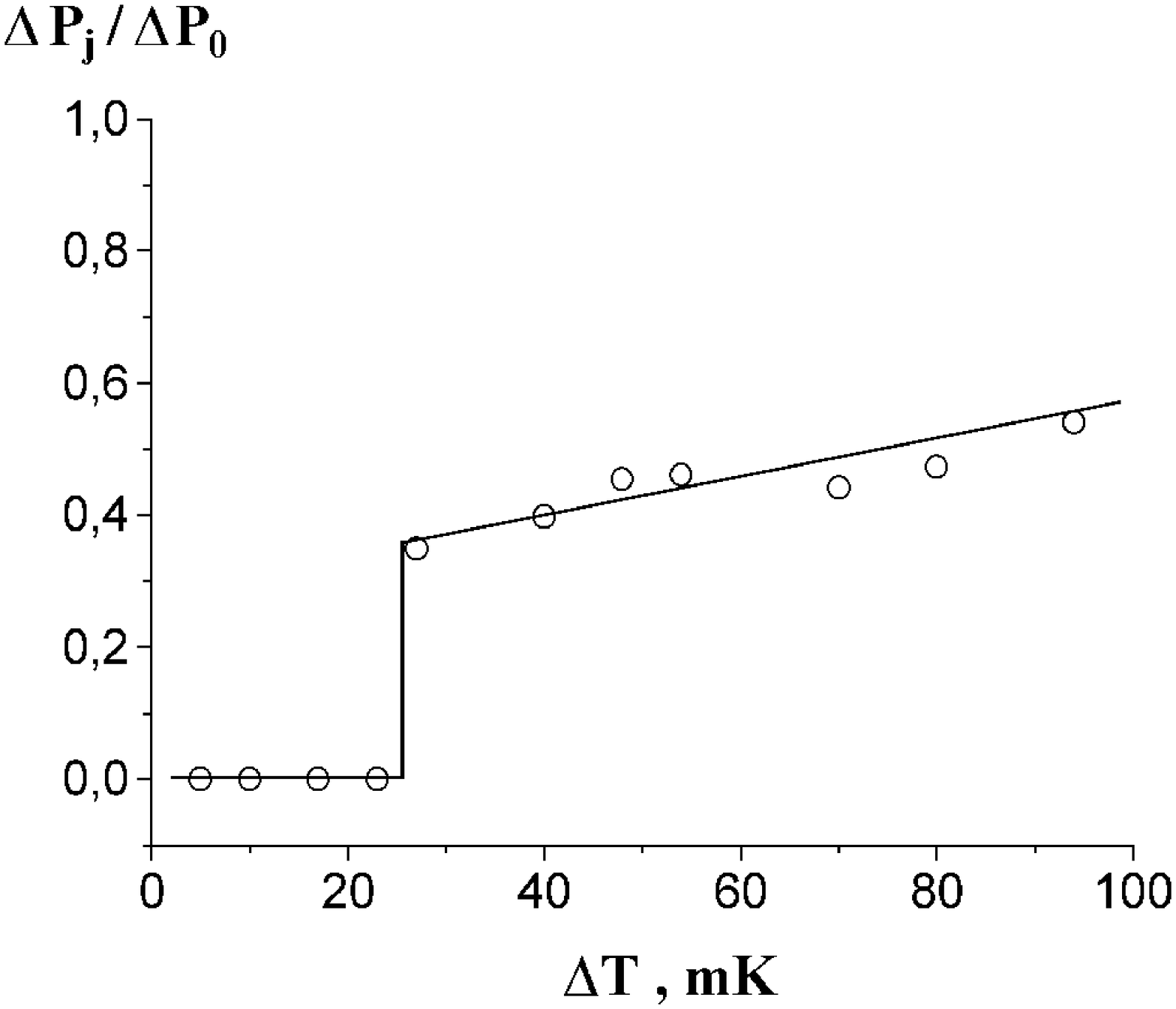}
\caption{ Relative pressure jump $\Delta P_j/\Delta P_0$ versus
difference between final and initial temperatures $\Delta T= T_f-T_i$.
Line is drawn to guide the eye.}
\label{fig2}
\end{center}
\end{figure}

\begin{figure}[t]
\begin{center}
\leavevmode
\includegraphics[scale=0.4,bb = 0 125 590 735,clip]{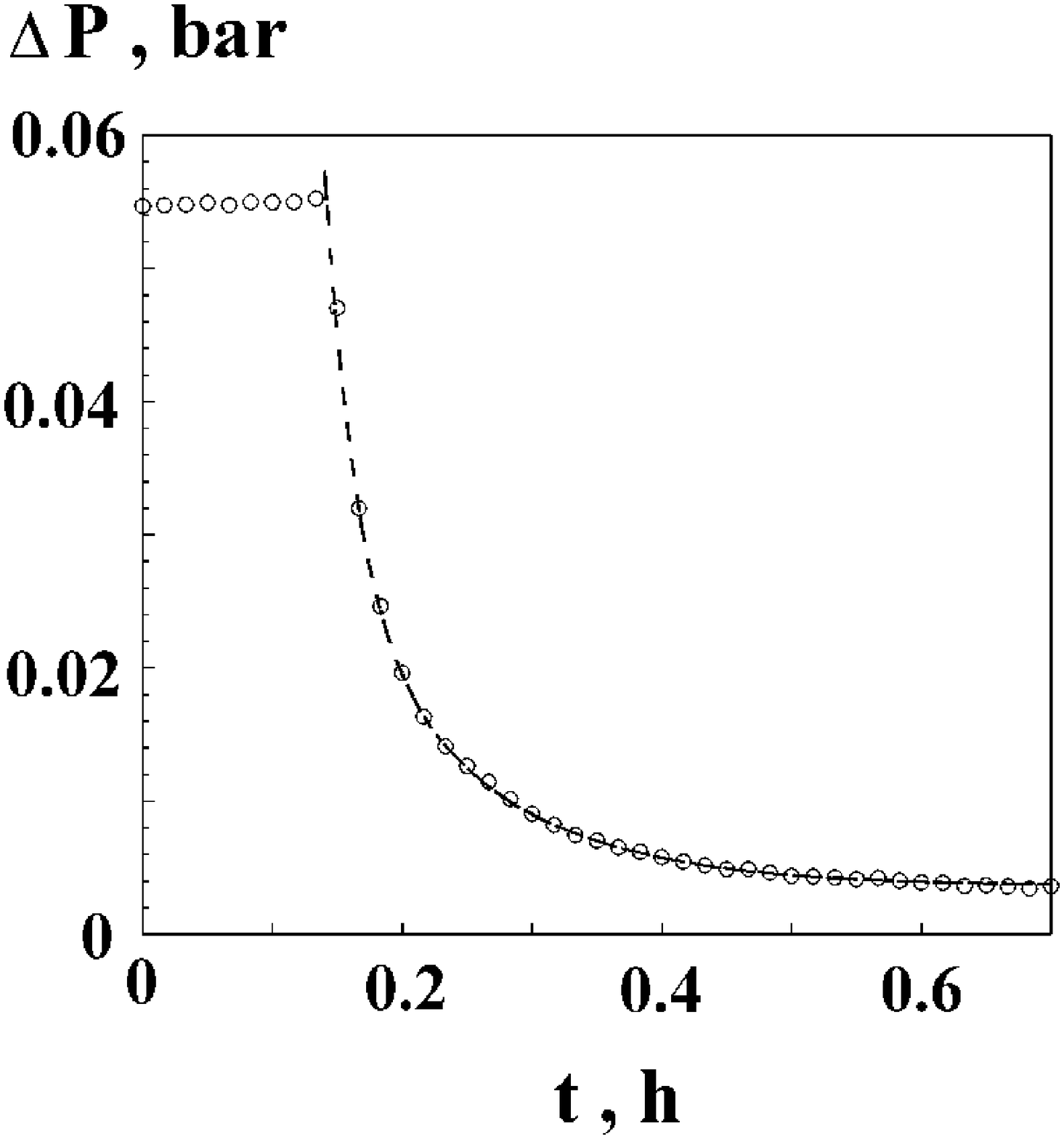}
\caption{ Time dependence of  pressure change in the sample during
$^3$He inclusion dissolution under warming from $T_i=150$ mK to $T_f=230$ mK.
The dashed line is fit to the equation: $\Delta P/\Delta P_0=
0.0472e^{-t/\tau_1}+0.2619e^{-t/\tau_2}$
 with $\tau_1\approx 85$ s and $\tau_2\approx 390$ s.}
\label{fig3}
\end{center}
\end{figure}

 The ratio of the pressure jump to the total pressure
change versus $\Delta T=T_f-T_i$ is of  threshold character
 (see Fig.\ref{fig2}).
  A pressure jump appears at some critical value $\Delta T_c$.
However, the obtained value $\Delta T_c \approx 25$ mK is likely to be
characteristic only of $T_i=103$ mK. As it has been found in
 two experiments carried
out with this sample that  at warming the crystal from 150 mK
to 230 mK the dissolution occurs without a pressure jump (see Fig.\ref{fig3})
in spite of the fact that the
 value of $\Delta T$ is essentially more than 25~mK.
From these two experiments it follows that
it makes no sense to consider a critical chemical potential difference since
 in this case it is several times larger than that corresponding
to the obtained $\Delta T_c$.


\section{MODEL OF THE DISSOLUTION PROCESS}
It is possible to explain qualitatively the observed effects,
assuming that the $^3$He inclusions dissolution process consists of
several stages. The essential feature of this model is the presence of a
strain layer in the vicinity of the inclusions boundary due to the great
difference between molar volumes of the inclusions and the matrix. The
elastic potential gradient appearance may result in the mismach of energy
levels in neighboring lattice sites and a
 restriction on impuritons quantum diffusion \cite{theory}.
 In particular, this circumstance is revealed
  at low temperatures when the effective diffusion coefficient, which
 characterizes the separation
 process, turns out much less than the quantum diffusion coefficient measured
in NMR experiments \cite{oursecond}.
 Therefore, quick enough substance transport near the droplets is
possible only after the elimination of the strain layer.

It can be realized at the first stage of the process, when the $^3$He
inclusions saturation with $^4$He atoms occurs.
 The diffusion penetration of $^4$He
inward the $^3$He inclusion, considered as a sphere of radius $R$, will
occur for the characteristic time
\begin{equation}
\tau_4 = \frac{R^2}{\pi^2 D },
\end{equation}
where the corresponding diffusion coefficient $D$ describes $^4$He atoms
 transport
in solid $^3$He by random tunnel jumps.  $D$
may be presented as \cite{theory}
\begin{equation}
D \sim J_{43}~a^2,
\end{equation}
where $a$ is the distance between the nearest neighbours and $J_{43}$ is the
exchange integral, which characterizes the tunneling frequency of $^4$He
 atoms  in $^3$He matrix. One would expect the magnitude of $J_{43}$ to be
close to the
tunneling frequency of $^3$He in $^4$He crystal $J_{34}$, which is precisely
determined in quantum diffusion experiments. Extrapolating well-known
values \cite{grigor} of $J_{34}$  to the molar volume $V\approx
24$ cm$^3$/mole, which
is characteristic of droplets, one can find $J_{34} \approx J_{43} \approx
2*10^7 $ s$^{-1}$. It gives the magnitude of $D\sim 10^{-8}$cm$^2$/s. Then,
for $R \sim 2 \mu$m (data of S.C.J. Kingsley et al.\cite{kingsley}) we obtain
 $\tau_4 \sim 0.1$ s.

Hence, the first stage of the dissolution is very fast. But the pressure
change $\Delta P$ at this stage must be very small because it is proportional
to the product of the concentration change of $^4$He in a droplet
(according to the
separation phase diagram \cite{balibar}) times the volume fraction occupied
 by the inclusions.
 The corresponding value of $\Delta P$ is estimated
to be not larger
than 2\% of the full pressure change $\Delta P_0$.

Nevertheless, this stage can strongly affect the droplets
dissolution process. When leaving a strained layer around droplets, $^4$He
atoms destroy it and make, therefore, possible free motion of the impuritons,
which determines the second stage of the process.
The characteristic time for this stage is determined by the velocity of
the $^3$He atoms egress from inclusions which can be estimated by analogy with
 evaporation of liquid. The number of atoms, evaporating from a unit of
surface per unit of time, is determined by \cite{frenkel}
\begin{equation}
\alpha \sim \vartheta N,
\end{equation}
where $\vartheta$ is the mean velocity of atoms in  vapor and $N$ is
 density of the atoms.

In our case $\vartheta \sim a \Delta _3 $ (see\cite{theory}) ($\Delta_3$ is
 impuritons
bandwidth in hcp phase). The density of $^3$He atoms in mixture $N_3$,
which is proportional to the concentration, serves as $N$.
\begin{equation}
N_3=N_m~x,
\end{equation}
where $N_m$ is  density of the atoms of the mixture.

 Then, it is possible to
find out  evaporation time for a droplet of radius $R$, neglecting the
difference in molar volumes of the inclusions and surrounding mixture
\begin{equation}
\tau \sim \frac{R}{3\Delta _3 a x },
\end{equation}
and assuming in Eq.(6) $R \sim 2\mu$m (data of S.C.J.Kingsley et
al.\cite{kingsley}),
 $\Delta _3 \sim 10^{7}$~s$^{-1}$ (from \cite{grigor})
and $x = 10^{-2}$ \% (from separation phase diagram\cite{balibar})
 gives $\tau \sim 1$~s. This stage turns out
 quick enough too.

The spherical layer, which is rich in $^3$He, is formed around the droplet
after this stage. For $x\sim 10^{-2}$ \% $^3$He the mean free path
of $^3$He impuritons in the mixture will be $\lambda_3\sim 10^2 a$
(from \cite{allen}). The formation of such a layer with the estimated thikness
of the order of several $\lambda_3$ ($\sim 10^{-5}$ cm) can provide the
value of the pressure jump $\Delta P_j$ observed in the experiment.
One can see from Fig.2 that $\Delta P_j\approx 0.4-0.5\Delta P_0$, where the
full pressure change $\Delta P_0$ is proportional to the concentration
change $\Delta x$ after heating the two-phase crystal of $\Delta T$.
According to the phase separation diagram \cite{balibar} $\Delta x
\approx 0.2 $ \% $^3$He for $\Delta T\approx \Delta T_c$. So, about 10 \%
of the $^3$He atoms in the crystal transfer from the inclusions to the matrix
and this leads up to the pressure decrease $\Delta P_0$. Therefore,
the observed jump $\Delta P_j$ can be provided by only about
 4 \% $^3$He atoms.

The propagation of $^3$He on a larger distance should occur by  diffusion
 due to the great probability of U-processes at impuritons collisions.
The characteristic time for this third stage of the dissolution will be
of the same order of magnitude as that obtained in other situations of the
 phase separation of solid $^3$He - $^4$He mixtures
 (see, for example, \cite{oursecond}).

The threshold effect in the proposed model takes place when the strain layer
around the $^3$He droplet preventing diffusion is completly eliminated
 as a result of the first stage, namely the dissolution of $^4$He.
 Since the quantity of dissolved $^4$He is uniquely determined by the
equilibrium concentration difference at the initial and at final
temperatures, the critical conditions correspond to such a difference of
temperatures $\Delta T_c $ at which the concentration difference $\Delta x$
provides the full elimination of the strain layer.

The equilibrium $^4$He concentration difference in the inclusion at the
final temperature $T_f \sim$ 130 mK, which corresponds to the critical
pressure
jump and the initial temperature $T_i\approx 103$ mK is about 0.2 \%.
 The spherical layer
of thickness $\delta R\simeq0.7\cdot 10^{-3}R \simeq1.4\cdot 10^{-7}$ cm must
be dissolved in the inclusion for providing such a value of the
concentration difference. The critical value $\Delta T_c$ is achieved by
dissolution of 3$\div$4 atom layers. It is a value reasonable enough value
for the layer thickness in which the quantum diffusion may be impeaded.
 It should be pointed out that  dissolution of $^4$He is
connected with decreasing density, and it provides some additional removal of
 strains.

The second necessary condition to observe the pressure jump is a impuritons
long mean
free path  in the mixture at the initial concentration
(temperature). The result of the above mentioned experiments with the sample
warmed from 150 mK testifies in favor of this confirmation. The equilibrium
concentration of $^3$He at 150 mK is about one order of magnitude more than
that at 100 mK. Accordingly, the mean free path of impuritons
is of the same times less, and the process of inclusions
 "evaporation" cannot cause the
perceptible pressure jump. Apparently it speeds up dissolving the
inclusions essentially and provides the effect comparable with the
following diffusion  as a result the dependence of $\Delta P(t)$ turns out
a sum of two exponents, which is illustrated in Fig.\ref{fig3}.

\section{CONCLUSION}
Thus, dissolving  droplets under heating from $T_i \simeq 100$ mK
 to $T_f=125\div 230$ mK can be considered to consist of three stages:
 $^{4}$He dissolution in bcc $^{3}$He inclusion, which removes
 the strain layer around the droplet;
formation of a rich $^{3}$He region around the inclusion;
diffusion dissolution of $^3$He.
The first and the second stages last for several seconds, but the third
 one lasts for about the time observed in other cases.

 The first stage of
 the process will not have the above indicated  concentration limitation and
can provide the perceptible pressure change under higher overheating,
when the sample is outside the phase separation region. It has been confirmed
 experimentally that in this case $\Delta P(t)$ dependence is described by
a sum of two exponents.

Unequal processes of the droplets growth and dissolution \cite{ourfirst} are
natural enough in the context of the considered model because the first of
them includes only the slowest stage. In doing so, the quantum diffusion
process turns out restricted at the expense of the strain layer near the
droplet boundary. In this connection it should be noted that the lack of the
sharp pressure jump in the experiment on dissolving the droplets, which
is illustrated in Fig.6. in \cite{ourfirst},
  is related with warming the sample much
 slower (as compared with the procedure used in the present work).

At  present  the experimental data which contradict  the
supposed dissolution model are obviously unknown. One will be able to say
with a fair degree of confidence about its reality only after the
verification of predictions made on its basis. Such predictions
may be the following:
  expected decrease in the pressure jump and  gradual disappearance of
this jump owing to reducing the mean free path of impuritons at
increasing $T_i$;
  increase in the value of $\Delta T_{c}$ for the liquid $^{3}$He
droplets  (at a lower pressure). It must cause increasing the
strain near the droplet boundary due to the density difference  enhancement.

Our experiments will be continued in this direction.
A theoretical treatment of the  questions considered would be desirable.

\section*{ACKNOWLEDGMENTS}
The discussions with A.F. Andreev, Yu.M. Kagan, and K.E. Nemchenko are
gratefully acknowledged.
This work has been supported in part by the International Program on Science
and Education (Grants QSU 082169 and QSU 082048).

\end{document}